# The Evolution Of The Digital Inheritance: Legal, Technical, And Practical Dimensions Of Cryptocurrency Transfer Through Succession In French-Inspired Legal Systems

Cristina Carata
*Department of Computing*
*Imperial College London*
London, United Kingdom
cc3619@ic.ac.uk

Ana-Luisa Chelaru
*Chamber of Public Notaries Bucharest*
Bucharest, Romania
luisa.chelaru@notariat-chelaru.ro

*Abstract*— **In recent years, cryptocurrencies have enjoyed increased popularity in all domains. Thus, in this context, it is important to understand how these digital assets can be transmitted, both legally and efficiently, in the event of the death of their owner. The present paper analyses the mechanisms of cryptocurrencies, analysing from a technical point of view aspects related to blockchain technology, virtual wallets or cryptographic keys, as well as various types of operations regarding this type of virtual currencies. The study also examines the legal aspects related to cryptocurrencies, with an emphasis on the diversity of their status in different global jurisdictions as well as the impact on inheritance planning. The case studies present tangible examples related to successions with cryptocurrencies as the main object, thus completing the exposition related to the main challenges faced by the heirs in the transfer process. In this way, this paper offers possible solutions and recommendations related to inheritance planning with cryptocurrencies as its main object, including the legal and fiscal aspects that must be taken into account when planning a digital succession.**

**Introduction**

*A. General context*

According to Romanian domestic legislation, a French-inspired legal system, virtual currencies are defined as "(...) a digital representation of value that is not issued or guaranteed by a central bank or public authority, is not necessarily linked to a legally established currency and does not have legal status of currency or money, but is accepted by natural or legal persons as a medium of exchange and can be transferred, stored and traded electronically"[1]. At the same time, digital assets like bitcoin can be broadly understood as digital representations of value, secured through cryptography, that can be transferred, stored or traded electronically through distributed ledger technology known as blockchain.

The first mention of the new type of digital currency – bitcoin – took place in 2008, with the publication of the whitepaper[2], the work detailing the concept of a decentralized virtual currency based on blockchain technology. The actual launch took place in 2009, with the first transaction made with the new digital asset, a moment that marks a real paradigm shift in the digital financial world. The Bitcoin white paper describes a decentralized, peer-to-peer electronic payment system that allows direct transactions between users without requiring a trusted intermediary. In recent years, the use of these digital assets has expanded considerably in various fields and they have found their applicability for a wide range of purposes, starting with investments or international money transfers and ending with decentralized financial applications. The rapid development has brought with it significant challenges for legislators at all levels, both at national and international levels, especially in terms of adapting laws to cope with the new possibilities offered by digital assets or creating new laws [1].

The private law aspects of digital currencies generate a number of legal dilemmas related to their rights, such as the right to possess, transfer or pledge. Traditional concepts used in private law are challenged by digital assets both because of their new and complex nature and through the cross-border situations they involve. Currently, there is no unified international regime in the field of digital assets. The potential norms of substantive international law in this area only cover some of the main aspects of digital currencies [1].

*B. Motivation and relevance*

Although the popularity of cryptocurrencies has grown exponentially in recent years, the issue of their transfer to heirs in the event of the owner's death remains insufficiently explored in literature. Compared to traditional financial assets, digital assets such as cryptocurrencies are held together in digital wallets that are protected by private keys. Access to this type of wallets is essential to manage and transfer digital currencies. Thus, cryptocurrencies can become inaccessible after the death of the owner, leading to significant financial

---

[1] Law no. 207 of July 21, 2021, amending and supplementing Law no. 286/2009 on the Criminal Code, and establishing measures to transpose Directive (EU) 2019/713 of the European Parliament and of the Council of April 17, 2019, on combating fraud and counterfeiting of non-cash payment instruments, as well as the Council Framework Decision 2001/413/JHA, published in the Official Gazette of Romania, Part I, no. 720 of July 22, 2021

[2] Within the meaning of Regulation (EU) 2023/1114 of the European Parliament and of the Council of May 31, 2023, on markets in crypto-assets, the whitepaper (translated as 'white paper for crypto-assets') is an informational document containing general details about the proposed project. The Bitcoin whitepaper is available at https://bitcoin.org/bitcoin.pdf [accessed July 21st, 2024]

losses, in the absence of a clear mechanism for the transfer of these private keys. The legal aspects of "digital" legacies are further complicated by both their decentralized nature and the pseudo-anonymity of cryptocurrencies. In the case of many jurisdictions, the current legislation does not explicitly cover cryptocurrencies in the context of succession, which favours the emergence of legal gaps and uncertainties. For this reason, the evolution of legal systems in the sense of integrating these new types of digital assets appears essential, thus ensuring the protection of the rights of the successors as well as the avoidance, as much as possible, of legal disputes. The safe and efficient transfer of cryptocurrencies requires technological solutions that allow successors to access digital wallets without compromising their security. The development of standardized practices and procedures for digital legacy management is also recommended, including recommendations for the secure storage and sharing of private keys.

This article aims to explore both the legal as well as the technical and practical aspects of digital currency inheritance in order to provide a comprehensive analysis of the challenges and possible solutions regarding their transfer. The main objective is thus to analyse and identify the legal and technical challenges associated with the transfer of digital currencies. In the context of the complexity and rapid evolution of blockchain technology and cryptocurrencies, the study aims to provide a set of recommendations to practitioners in the field to improve existing legislation and implement effective technological solutions.

The paper is structured as follows: Section I explores the mechanisms of cryptocurrencies, referring in detail to the blockchain technology underlying these types of digital currencies, as well as essential aspects related to their trading, such as wallets, cryptographic keys or exchange platforms. Section II addresses the cryptocurrency inheritance from a legal and practical perspective, starting with an analysis of the legal status in various jurisdictions and ending with estate planning. Section III presents the challenges encountered in practice, with a special emphasis on case studies that support the understanding of the problem. Section IV develops the fiscal and legislative implications in the inheritance of cryptocurrencies while the Section V concludes the article.

### I. How Cryptocurrencies Work

For a better understanding of cryptocurrency inheritance, the next section explores the way how these digital assets function as well as the ecosystem evolving around which is comprised of different technical elements such as blockchain technology, digital wallets or mining.

#### A. Blockchain technology

The blockchain concept was first exposed in 2008, with the publication of the Bitcoin white paper, signed with the pseudonym Satoshi Nakamoto, used to protect the author's identity. This whitepaper marked a turning point, technologically speaking, by introducing a new way of managing and transferring digital assets. The system described in the paper was based on a peer-to-peer electronic network, that eliminated the need for a traditional intermediary such as banks and allowed transactions to be carried out directly between the users of the network. Although not obvious at that moment, the blockchain technology innovation not only fundamentally changed the classical financial systems, but also paved the way for multiple further applications, both within cryptocurrencies and in other areas. This type of decentralized approach theoretically reduces the risk of fraud and eliminates the need for a trusted central authority, an essential feature of cryptocurrencies [2]. Blockchain technology is vital to the functioning of cryptocurrencies and plays an essential role in the context of digital heritage. In short, blockchain can be defined as a digital ledger of data distributed among all participants, which records all transactions made with cryptocurrencies and ensures the integrity and transparency of these transactions.

A blockchain data ledger is composed of a series of "blocks" (containing data) linked together in chronological order, each block containing a set of verified transactions and a cryptographic hash - a unique digital fingerprint - of the previous block. The hash is a unique identifier generated by applying a cryptographic function to the data in the previous block, thus ensuring the integrity of the entire chain as a whole. By using this mechanism modifying a previous block is extremely difficult, as it would require recalculating the hashes for all subsequent blocks, an action that would be quickly detected by the rest of the nodes/participants in the network [3]. Thus, the blockchain ensures high transparency and security, being a tool for recording and verifying financial transactions and more.

Blockchain technology works through a decentralized network of nodes – where each node is a computer. Nodes maintain and validate this type of network by having a complete copy of the data ledger and actively participate in the transaction validation process. In short, with the initiation of a new transaction, the entire network receives it and it is recorded in the data ledger. The next step is represented by nodes that validate the transaction through technical cryptographic processes while verifying digital signatures and ensuring that the sender has sufficient funds available. Validated transactions are grouped into blocks and added to the blockchain [4]. The process is called "mining". This decentralized validation process gives the blockchain system a high degree of security.

Mining is the process by which nodes (or miners) compete to add a new block to the blockchain. Miners solve complex mathematical problems to find a cryptographic code (known as a hash) that meets certain criteria (for example, starts with a certain number of zeros). The first miner to solve the mathematical problem adds a new block to the blockchain network and receives a reward in the form of cryptocurrencies. This type of consensus mechanism is known as "Proof of Work" (PoW). Thus, the security and integrity of the network is ensured in a way that makes it difficult to modify already recorded transactions. This type of consensus mechanism is essential in preventing attacks and various forms of network manipulation for the simple reason that it requires a high computational effort to validate transactions.

Concluding the above, the following characteristics of blockchain technology can be distinguished:

1) It is a digital data register, visible simultaneously by all network participants;

2) Records all digital currency transactions;

3) Information is grouped into "blocks". Each block contains a set of verified transactions (e.g.: who sent money to whom and how much);

4) Each new block is linked to the previous block by a "hash," a unique code that works like a fingerprint. This hash ensures that the blocks are in chronological order and their integrity is maintained;

5) Blockchain works on a network of computers called "nodes". Each node owns a complete copy of the blockchain;

6) When a new transaction is created, it is broadcast to the entire network;

7) Once added to the blockchain, a block cannot be modified without modifying all subsequent blocks, which is very difficult and would be detected immediately. This ensures the security and transparency of the entire system.

## B. *Cryptographic wallets and keys; transactions and confirmations*

To understand how cryptocurrencies are traded, an analysis of the basic elements that make up this ecosystem is necessary, in this case digital wallets, cryptographic keys and exchange platforms.

Digital wallets for cryptocurrencies are fundamental to the use of blockchain and digital currencies due to the fact that every user who intends to use the blockchain platform for any transaction must use such a tool. Unlike traditional wallets, cryptocurrencies are not actually stored in this type of wallet. Cryptocurrencies are not stored in a single location and do not physically exist anywhere; they only exist as transaction data stored on the blockchain. Wallets allow the user to create an account, i.e., a pair of private key and public key (which will be discussed later) and store them in such a wallet. To make any transaction on the blockchain, a user must sign ownership of coins to their wallet address. There is no actual exchange of actual coins, but an exchange of transaction data values created on the blockchain, thus making a change in the balance of the user's digital wallet [5].

Cryptocurrency wallets can be classified into several categories:

1) "Desktop" wallets are software applications that are installed on a computer and give their users full control over their funds. This type of wallets offers a high level of security due to the control over the keys, but they become vulnerable when the device they are installed on is compromised (for example, infected with a virus);

2) Web wallets are software applications based on cloud storage. This way, they are accessible from any location using a web browser. In the case of these types of wallets, the private keys are managed by third parties, thus making them susceptible to online hacking risks;

3) Mobile wallets are smartphone apps that allow users to manage their cryptocurrencies. Although they are considered to be more secure than the previously mentioned online wallets, they can still be compromised if the phone is lost or infected with a virus. They are easy to use and convenient for payments in commercial environments;

4) Hardware or "offline" wallets are physical devices, such as USB sticks, that store private keys, disconnected from the Internet. They are considered the most secure ways to store cryptocurrencies, because by not having access to the internet, they protect your private keys from online threats. They are ideal for long-term storage and large amounts of cryptocurrency;

5) Cold wallets refer to any type of wallet that remains disconnected from the Internet, including hardware and paper wallets. These wallets are very secure because they are not connected to the internet, being immune to online hacking attempts. They are best suited for long-term storage of a large amount of cryptocurrency;

6) Hot wallets are wallets that are always connected to the internet, facilitating frequent transactions. This type of wallets are convenient for daily use but are more vulnerable to online threats (compared to cold wallets, for example);

7) Paper wallets involve printing the public and private keys of a cryptocurrency on paper. These keys can also be stored as QR codes for easier use. Although it offers a high level of security, there is a risk of the physical document being lost or stolen. They are suitable for safe and long-term storage of cryptocurrencies, although less convenient for daily transactions.

When choosing a cryptocurrency wallet, one should consider according to security and personal needs. Generally, the specialized literature recommends a combined use between an Internet-based wallet for daily transactions and a hardware wallet for long-term storage. Each type of wallet has unique features and security levels, so it is essential that the selection is made according to individual requirements and usage pattern [5].

Cryptocurrencies can offer their holders an increased degree of privacy compared to traditional currencies, mainly due to their different approaches to recording ownership and validating transactions. Technically, cryptocurrencies work with the digital keys of their holders. These keys are generated in pairs: each pair contains a public and a private key; the keys are created and stored by the aforementioned digital wallets. As the name suggests, the public key is publicly available and is used to generate an address, where ownership of the cryptocurrency is registered. Compared to the traditional financial system, the public key can be likened to the bank account number: it is public and can be transmitted to other people for the purpose of transactions, without the risk of endangering the existing funds in the account. The private key belongs strictly to the holder and is never disclosed to preserve the integrity of available funds. Compared to the traditional financial system, the private key can be likened to the PIN code. To authorize a cryptocurrency transaction, a digital signature is generated with the private key to show ownership of the cryptocurrency associated with the public key. Compared to this key-based cryptographic model of digital currencies, traditional currencies are associated with their owners' personal information instead of such cryptographic keys. Unless cash is used, the ownership of traditional coins and the personal information of the owners is permanently known by the financial institutions (banks). Classic transactions through financial institutions are also authorized with this personal information [6].

Exchanges are a technical platform where users can buy and sell cryptocurrencies. Most of the exchange platforms in the current market only offer trading services between cryptocurrencies, but recently there has been an increase in platforms that offer exchanges between digital currencies and fiat currencies (for example, US dollar or Euro). Similar to the stock market, users also use cryptocurrency exchange platforms to benefit from cryptocurrency price changes. There are three types of exchange platforms: centralized platforms (CEX) that are governed by a company or organization, decentralized platforms (DEX) that provide an automated process, independent of any other entity, for transactions, and hybrid platforms that combine both elements of above [7].

Concluding, a digital currency transaction follows the next steps:

1) Creating a digital wallet;

2) Initiating a transaction to another user (for example, user A wants to send user B the amount of 1 bitcoin) – the recipient's wallet address is needed;

3) With the help of the private key, the transaction is signed – through this action, a so-called digital signature is created that proves the right to control the cryptocurrency and thus authorizes the transaction;

4) The transaction thus signed is registered in the network for the visibility of all participants and for its validation;

5) The participants in the blockchain network (the nodes mentioned in Section I.A validate the transaction by checking the balance and by checking the correctness of the digital signature – from a technical point of view, this validation is carried out through a process called mining, which we will return to in the next section;

6) After the transaction has been verified, it is included in a block that is added to the blockchain – this process ensures that the transaction is permanent, cannot be changed and is visible to all users of the network.

As an additional note, it should be remembered that there are several ways to acquire cryptocurrencies, the most popular being mining (creating them by participating in the block production process) and purchasing through exchanges (obtaining through via trading platforms). In short, without going into technical details, the process called mining - or the extraction of cryptocurrencies - is the process by which specialized computers solve complex mathematical problems to add new blocks to the blockchain. These computers (or "miners") use the computing power in order to perform the necessary calculations and in return receive a reward in cryptocurrency [8]. This is the "original" method by which all circulating cryptocurrencies are created. The second method, purchasing via trading platforms, involves creating accounts on dedicated platforms (such as Binance or Coinbase). After depositing fiat funds, users can buy cryptocurrencies at the market price and ownership is transferred once the transaction is completed, with the cryptocurrencies stored in digital wallets [1].

II. INHERITANCE INVOLVING CRYPTOCURRENCIES. LEGAL AND PRACTICAL ASPECTS

A. *The legal nature of cryptocurrencies and their legal status*

Given the growing frequency of legal cases and challenges associated with cryptocurrencies, it has become essential to prioritize addressing the legal status of these digital assets. Although a unified regulatory framework has yet to be established globally, recent efforts to provide legal classification and regulation for such assets have culminated in the European Regulation known as MiCA (Markets in Crypto-Assets). This regulation sets forth the legal framework applicable to EU Member States, which will be explored in further detail below.

The legal nature of cryptocurrencies varies in determination in different global legislations. In the vast majority of cases, they not considered as currency in the classical sense of the term, since they are not centralized and do not represent a compulsory means of payment. At the same time, the literature [9] highlights the main differences between traditional electronic currencies and cryptocurrencies, their legal and financial nature being determined by a number of specific characteristics:

    1) There is no central cryptocurrency issuing authority or central cryptocurrency administrator;

    2) Payments in a cryptocurrency system are pseudo-anonymous. The addresses from which money flows in and out are strings of numbers and digits with no specific additional owner information being recorded in the system;

    3) They circulate in a decentralized and accessible system. Any person can connect to the blockchain network with their own digital wallet and thus connected, verify the transaction history;

    4) Payments are fully transparent since information about a transaction remains recorded in the system and cannot be erased;

    5) There is no single central authority with control abilities. Due to the fact that the blockchain network is a user-to-user database, with certain exceptions due to cases of necessity, no user can block a specific account. Transactions in the system are, in principle, irrevocable;

    6) Unprecedented protection. The longer the chain, the harder it is to crack it or forge it. To date the Bitcoin network has surpassed the processing power of the world's largest supercomputers [10].

While there are numerous similarities between cryptocurrencies and traditional money, the most notable distinction lies in the entity responsible for issuing them. Both forms of currency share certain fundamental characteristics, as outlined below [10]:

    1) Bitcoin is a decentralized system that generates cryptocurrency with a specific financial and legal nature;

    2) Cryptocurrencies can be exchanged for other goods or services by agreement of the parties;

3) Cryptocurrency is characterized by a high degree of liquidity, just like fiat money, as long as it can be quickly exchanged for cash without losing significant value;

4) Cryptocurrency is divisible. For example, one Bitcoin is subdivided into 100,000,000 units arbitrarily called Satoshi. So, you can pay a certain amount, get change, etc.;

5) Cryptocurrency is portable. A transaction can be made in minutes from anywhere in the world;

6) Cryptocurrency has no value in itself. Its value is determined by the number of goods and services that can be obtained in exchange for a given number of cryptocurrencies;

7) The network through which it flows - as mentioned above, the blockchain shows all transactions made. Any user can verify a transaction and the location where it was made;

8) Cryptocurrency, unlike fiat money, has no centralized issuer. The coins are generated by a separate group of users who are called miners.

Although opinions on the legal nature of these cryptocurrencies are divided, they are considered by the majority in literature and doctrine as goods that can constitute the subject of property rights, with all the consequences that this entails [11]. Cryptocurrencies can be qualified as objects of property rights, as long as they have an economic value, present a certain financial interest and thus, they can be the subject of transactions, exchanges, or other means of transfer of property rights such as inheritance. The legal nature of cryptocurrencies is similar to the concept of a product [9].

*Legal status*

The new European legislation that has recently entered into force, namely the European MiCA Regulation[3], through the provisions of Art. 3 para. (1) point 5, defines cryptocurrencies or cryptoassets as "a digital representation of a value or a right that can be transferred and stored electronically, using distributed ledger or similar technology".

Before MiCA regulation, cryptocurrency legislation was a patchwork of national laws, with EU countries taking different approaches to these assets. For example, in 2014 Spain recognized Bitcoin as an official payment system. In Italy [12], the first regulation of cryptocurrencies appeared in 2017 and they have been defined as electronic currency. According to art. 810 of the Italian Civil Code they are categorized by their legal nature as goods, being susceptible to form the object of property rights. In France, in the context of its alignment with European legislation, the regulation of cryptocurrencies has evolved significantly in recent years, in line with the growing interest in cryptoassets, as well as innovation, investor protection and market integrity, which are obviously also the aims of the European Regulation. The most comprehensive definition under French domestic law is provided by the Pact Law[4]. According to it, digital assets are "digital representations of value that are not issued by a central bank or public authority, are not necessarily attached to a legal tender and are accepted by natural or legal persons as a means of exchange and can be transferred, stored or traded electronically". Malta has declared itself as a "Blockchain Island", being one of the first and few countries in the European Union to create the domestic legal framework by adopting the Digital Innovation Authority Act, the Innovative Technology Agreements and Services Act and the Virtual Financial Assets Act, which aim to regulate cryptocurrencies and Blockchain transactions [13, 14, 15]. Although not a member state of the EU, Switzerland has adopted a very favourable tax legislation on cryptocurrencies and the first cryptocurrency exchange agency, ECUREX Gmbh [16] also emerged in this country, complying with all the rules required by Swiss banking legislation. In addition, cryptocurrency transactions have been exempted from VAT, a clear proof that these transactions are considered as payment methods and not as methods of transfer of goods.

The emergence of the MiCA Regulation was necessary in order to create a specific harmonized framework at EU level for crypto-asset markets, containing specific rules for cryptocurrencies and related services and activities that were not covered by Union legislation on non-financial services. As can be seen, until its emergence, there were no uniform rules in the Member States of the European Union for the provision of services related to crypto-assets, their trading, their exchange for funds or other crypto-assets, or for their administration to customers, except for those relating to anti-money laundering. The lack of an overarching EU regulatory framework for crypto-asset markets was leading to "regulatory fragmentation" which distorted competition in internal markets, while at the same time there were a multitude of risks to which cryptocurrency holders were exposed, particularly in areas not covered by consumer protection legislation [17].

Although it covers an area not yet fully legislated, the MiCA Regulation is intended to be the global point of reference [18] and covers natural and legal persons as well as certain undertakings that are involved in the issuance, public offer and admission to trading of cryptoassets or those providing cryptoasset services in the European Union [17]. At the same time, it introduces a general framework aimed at ensuring market integrity and smooth market functioning, promoting innovation and ensuring consumer protection within the EU financial landscape [19,34]. MiCA classifies cryptoassets into three types, which are differentiated from each other, each of which should be subject to specific requirements, depending on the risks they entail [17], the

---

[3] Regulation (EU) No 2023/1114 of the European Parliament and of the Council of May 31, 2023, on Markets in Crypto-Assets (MiCA), amends Regulations (EU) No 1093/2010 and (EU) No 1095/2010, as well as Directives 2013/36/EU and (EU) 2019/1937. A provisional agreement between the European Parliament and the Council on the MiCA proposal was reached in June 2022. MiCA was subsequently ratified by the European Parliament on April 20, 2023, marking the world's first comprehensive regulatory framework for crypto markets. This regulation establishes clear standards and guidelines for market participants, with the primary objectives of ensuring consumer protection and maintaining market integrity. The implementation phase of the Regulation is scheduled to conclude in December 2024, followed by an 18-month transition period ending in July 2025. By that time, all entities providing crypto-asset services must fully comply with the new EU regulatory standards. For further information, refer to the European Securities and Markets Authority's statement on MiCA supervisory convergence, available at https://www.esma.europa.eu/sites/default/files/2023-10/ESMA74-449133380-441_Statement_on_MiCA_Supervisory_Convergence.pdf (accessed July 21, 2024).

[4] Pact Law (Loi Pacte - Plan d'Action pour la Croissance et la Transformation des Entreprises) was enacted in France on May 22, 2019, with the objective of fostering business growth and transformation. The law seeks to simplify the regulatory environment, promote entrepreneurship, and support the development of businesses, particularly small and medium-sized enterprises (SMEs). A key aspect of the law includes provisions for modernizing the French economy, facilitating innovation, and addressing emerging areas such as crypto-assets and cryptocurrencies. For further details, see *Cryptoactifs, cryptomonnaies: comment s'y retrouver*, available at economie.gouv.fr, consulted on July 21, 2024

general framework also includes complementing existing financial legislation.

*Comparison with other legislative regimes*

A comparison [19] of the legal regulation of cryptocurrencies in the European Union, in the United States and in the United Kingdom reveals the different approaches, each characterized by different legislative philosophies, different legal systems, market dynamics and policy objectives.

In the United States, the legislative context is characterized by an amalgam of federal and state laws, and there is no unitary legislation covering this type of asset. Cryptocurrencies that qualify as securities under the Howey Test [20], fall under the purview of the Securities and Exchange Commission (SEC -Securities and Exchange Commission) [21]. At the same time, the Commodity Futures Trading Commission (CFTC) oversees Bitcoin and Ethereum coins that are classified as commodities [22] and the Financial Crimes Enforcement Network (FinCEN) enforces money laundering and consumer protection regulations, requiring cryptocurrency *exchanges* to register as entities offering monetary services [23].

The UK has taken a more consolidated approach to the regulation of cryptocurrencies since Brexit, with the FCA (Financial Conduct Authority) playing a central role in this area. The FCA requires all cryptocurrency business to be registered and to comply with anti-money laundering and know-your-customer regulations, while also creating the framework for resale under certain conditions. The Financial Service and Market Bill (FSMB) seeks to create more competitive financial services post-Brexit, with one of its aims being to integrate the movement of cryptocurrencies into the existing payments system, allowing for a flexible regulatory environment. Due to the fact that it is not intended to create a new circulation infrastructure for cryptocurrencies, this legal rule appears to be in contradiction with the MiCA Regulation [24].

The MiCA Regulation provides a single regulatory framework for all Member States, reducing the previously existing regulatory fragmentation seen in the US and the UK. The scope of MiCA covers a wide range of crypto-assets, whereas in the US they are considered securities, commodities or currencies. The UK approach is in between the two, focusing on consumer protection and upholding market integrity. MiCA establishes a clear supervisory structure, involving national authorities and EU-level bodies - the European Banking Authority (EBA) and the European Securities and Markets Authority (ESMA). On the other hand, the US uses multiple federal and state agencies, while the UK has a single supervisory authority, the FCA [19].

In conclusion, the introduction of the MiCA Regulation is a significant step forward in creating a unified regulatory environment for cryptocurrencies in the European Union, while the US and the UK continue to have their own fragmented legal regimes. The harmonized approach of the Regulation may also serve as a future model for other countries that would like to regulate the cryptocurrency market, with the overarching goal being to balance innovation with consumer protection and market integrity.

*B. Inheritance planning*

In recent times, the concept of digital inheritance has become increasingly important. In this context, part of the purpose of this article is to find answers to relevant questions such as: what happens to a person's cryptocurrencies post-mortem? Can they be transferred by inheritance? If yes, what is the actual procedure for implementation? How can legal heirs determine whether the deceased owned cryptocurrencies, and what steps must they follow to legally gain access to them? What legal implications may their transfer by inheritance have and what are the applicable tax rules?

As stated in the introductory part of this article, digital assets refer to any property that exists in digital form, whether it be email addresses, social media accounts, bank accounts, photos, videos, music, NFTs or digital currencies. In the past, in the French-inspired legal system, these types of assets were considered to be intangible and therefore not subject to succession. Subsequently, over time, a classification between these assets became obvious due to problems that arose in practice, and as a result, even service providers such as Facebook, LinkedIn or YouTube started to have contractual clauses on the fate of data after the death of the user. Although there is no specific unitary legislation on inheritance either, various countries have begun to legislate and recognize the possibility of cryptocurrencies being transferred by inheritance, as they are considered as assets that can be transferred both inter-vivos and mortis-causa. Due to the specific characteristics of these cryptocurrencies, their post-mortem transmission entails a number of issues which we will address below.

Firstly, there is the issue of decentralization. Comparing them to already known institutions, the cryptocurrencies and the wallet in which they are contained could be compared, by analogy, to the amounts in a bank account or to known stocks/bonds/ shares financial instruments. The problem is that, unlike these, when it comes to cryptocurrencies, there is no single central authority from which heirs can obtain information, (except for exchange platforms) authority/institution that has a centralized record of all cryptocurrencies issued and that can produce, on request, information on what, how many and if a person holds cryptocurrencies, and if held, what kind of cryptocurrencies they are.

Secondly, there is the issue of pseudo-anonymity and privacy. Cryptocurrencies are held in wallets whose owner is pseudo-anonymous/confidential, with only the numeric strings assigned to the wallets appearing in the blockchain network, not the identification of the wallet owner. The wallet can be owned by anyone in possession of the "wallet number" and access keys. Thus, the heirs have no way of knowing, and no way of finding out, whether their author has held cryptocurrencies, unless he declares them or at least mentions that he has held them during his lifetime.

Thirdly, technically, in order to acquire possession of these cryptocurrencies and to access a virtual wallet, a person, in our situation the heirs, must have the access keys. How will the transfer, after death, of these access keys to the heirs be handled?

Lastly, how will the inheritance transfer of such assets be evaluated and subject to taxation?

A study published 2020 [25] estimates that 89% of cryptocurrency investors are concerned about the fate of their cryptocurrency holdings after their death, while only 23% have a documented plan in place. Of the latter, only 7% have drawn up a legacy that includes digital currencies. At the same time, it is estimated that globally an estimated 20% of all bitcoins are lost and unrecoverable due to the death of the bitcoin owner and the lack of transmission of irrecoverable access keys. In this context and for these reasons, when talking about the succession of cryptocurrencies, a very strong emphasis on succession planning should be the first and foremost.

All cryptocurrency holders should be aware, or receive legal counsel in this matter, that it is essential to inform their heirs about their cryptocurrency ownership. This should be done through a will or another legally recognized method, specifying the location of the assets, such as providing the wallet's QR code or numeric key. As previously mentioned, due to the fact that the holders are pseudo-anonymous and that it is a decentralized system, the heirs do not have a central authority to whom they can turn to find out information about these holdings.

### III. APPROACHES AND CHALLENGES ENCOUNTERED IN PRACTICE. CASE STUDY

#### A. Cryptocurrency transfer methods by inheritance

In this regard, it is important to differentiate between various forms of holding arrangements. If cryptocurrencies are stored in electronic wallets, the approach suggested earlier, such as utilizing wills or other legal instruments, is applicable. Conversely, if funds are stored in certain *exchange* platforms such as Coinbase, Kraken, or Binance, these platforms may possess certain information about the cryptocurrency owners. In this hypothesis, provided that the heirs have prior knowledge of the deceased's cryptocurrency ownership on these platforms, these entities could potentially release the necessary information required to facilitate the inheritance process.

Secondly, in this matter, because it concerns movable property *(chattel),* the possessor is presumed to be the owner (possession equals ownership) [26]. Thus, in order to have effective access to the cryptocurrency in these wallets, heirs must obtain possession of the access keys. If a person acquires these keys, he/she automatically becomes the owner and user of that wallet and the funds in it. This is currently the biggest estate planning challenge for the transfer of cryptocurrencies. What is the best and most secure way to transfer these keys? Inheritance planning should also provide very clearly this method of transfer.

In some opinions, the transfer of keys should be made according to the type of wallet in which they are held. If it is an exchange platform such as those mentioned above, the heir can use the ways known so far to prove his or her heirship and be granted access to these accounts. However, if it is about digital wallets, the cryptocurrency holder should consider, when planning his or her succession, that he or she should leave to the heirs all the information about: what kind of wallet is being used, where it is installed, what kind of cryptocurrency is in the wallet and where the heir can find the access keys or the phrase, the words to recover these access keys in case of their loss. With regard to hard-drive wallets, the author must mention where the devices that represent the wallets (e.g., USB) and the passwords can be found.

Another perspective takes into account three categories that are most common in practice:

1) The traditional method. This involves the use of a will indicating the existence of the wallet and the access keys. The disadvantages of this method are that, due to certain circumstances, the anonymity specific to this type of holding is lost and there is also the possibility that the access keys may be lost forever. In this matter it is suggested that the will should still only contain the fact that the wallet and the public key exist, the private key should be kept separately;

2) The technological method, which involves various default transfer methods or programs that are in these wallets, certain automated transfer systems, based on smart contracts and proof of death of the holder. One of the most advanced methods of asset transfer involves the use of an electronic safe powered by blockchain technology. The owner also leaves instructions on the transfer of data to heirs, which can be triggered, for example, by submitting a death certificate. Apparently, even this technology cannot guarantee complete data protection. Many exchange platforms offer services to transfer a digital wallet to heirs after they provide proof of death and that they are the heirs;

3) The mixed method, includes inheriting the digital wallets in both physical and electronic format. Access keys can be left in places such as bank safe deposit boxes or warehouses, and their location can be indicated in the text of a will. [26]

It seems that at the moment, the most used method is the mixed method, which is considered to be the safest way of transfer.

#### B. Case studies

In order to emphasize the importance of the above, we present below two real-life situations regarding cryptocurrency succession, one that ended in failure, and one that successfully accomplished its purpose so that the cryptoassets were transferred to the heirs.

*The Quadriga CX case - Gerald Cotten's legacy*

Gerald Cotten was the founder and director of one of Canada's largest cryptocurrency exchange platforms called QuadrigaCX [27]. He passed away in December 2018, at the age of 30, while traveling with his new wife to India. His sudden death created a huge problem over access to the platform's client (investor) funds, which were valued at around 190 million Canadian dollars. In fact, he was the only person who had access to the keys needed to access the digital wallets containing these coins at QuadrigaCX. Cotten left no instructions to retrieve those keys, so that access to customer funds was not possible, making it impossible for them to recover the money.

After his death, his wife informed the public and the authorities about the situation. An investigation was launched due to the huge amount of money on the platform, with approximately 115,000 people having suffered losses, and the fact that the sudden death was viewed with suspicion, with most investors believing that they were dealing with fraud rather than a death-related situation. Financial security experts and auditing firms were hired to try to locate and access the funds. All of Cotten's electronic devices, documents and possible sources of the keys were analysed.

During investigations, some of the funds were recovered, but much of the cryptocurrency held by QuadrigaCX remained inaccessible, eventually leading authorities to conclude that it was all a Ponzi scheme, with Cotten defrauding customers by crediting their accounts with fictitious amounts, or investing them in other currencies that were later devalued [28]. On the other hand, it seems that due to the inactivity found in Quadriga's electronic wallets, it is believed that he may indeed be deceased.

This case highlights the risks associated with the centralization of control over cryptocurrencies but also the lack of a contingency plan. Proper planning and security measures are essential to ensure the safe and efficient transfer of cryptoassets in the event of an unexpected event.

*Hal Finney's legacy*

An example at the opposite direction is the one represented by Hal Finney, computer scientist and cryptographer, one of the pioneers in the field of cryptocurrencies as well as the first person to receive Bitcoins, marking the first transaction in history, on January 12nd, 2009. He died in the August 2014, leaving a considerable legacy in Bitcoin, a sum raised due to his early involvement in the development of this digital asset.

Finney was diagnosed with a degenerative disease that allowed him to plan his inheritance in detail, including the cryptocurrencies he owned, until the moment of his death. Thus, being aware of his medical condition, Finney submitted all the necessary diligence so that his wife, Fran Finney, and other heirs will be able to access the funds, especially those in cryptocurrencies. Apparently, he left the necessary documentation, private keys and the procedure for accessing electronic wallets in a safe. [29,30].

After his death, the heirs followed the instructions left and were able to access and take possession of the cryptocurrencies. In conclusion, due to the planning and precautions taken by Hal Finney, the heirs were able to access cryptocurrencies without major problems. The case highlights the importance of detailed planning and preparation for inheriting crypto assets.

## IV. TAX AND LEGISLATIVE IMPLICATIONS IN CRYPTOCURRENCY INHERITANCE

### A. Tax implications

Like any inheritable asset, cryptocurrencies are subject to inheritance tax in most countries in the world. First and foremost, as they are considered property in the legal system, property over which a right of ownership is constituted, they can be transferred between two or more persons, including *mortis causa*, by succession. Cryptocurrencies form part of the estate of the deceased and can be bequeathed to descendants or persons designated in a will [31]. In France, since 2014, the central administration has introduced in its doctrine the principle that cryptocurrencies can be passed on to heirs in order to include crypto-assets in the field of assets to be declared in a succession [32]. Secondly, as far as taxation is concerned, they are subject to inheritance tax in accordance with the laws of each state in which they are inherited. They will be taken into account and treated in the estate in the same way as other assets such as money, shares, real estate, etc. For example, in the UK, bitcoins or other digital assets are considered property for inheritance tax purposes. With certain exceptions, they are subject to inheritance taxes of up to 40% that apply to estates in that country, and in Germany, inheritance taxes range between 7% and 50% of the value of the inherited assets. Bitcoin as well as other cryptocurrencies are included in the total estate left to descendants. In the US, descendants must pay inheritance tax on cryptocurrencies according to the total estate they inherit, and in Japan, crypto-assets are treated in the same way as any other assets (money, securities, real estate, etc.) from an inheritance point of view. In France, similarly, cryptoassets are subject to inheritance tax that applies to all inherited assets, with taxes ranging from 5% to 45%, depending on the value of the estate.

As far as taxation is concerned, the biggest issue that has been raised in this area is the risk posed by the volatility of these currencies. Who will bear it, given that their value is constantly changing and is not fixed in relation to an official rate? And, at the same time, what is the point in time against which their value is fixed? In this respect, all the doctrine and literature have concluded that the volatility risk will be borne by the heirs and that the value of the cryptocurrencies to be taken into account in the succession is the value they had at the date of death of the person leaving them. Given that the right of ownership is acquired retroactively, from the date of death, this is also the point at which the valuation will be determined. Their valuation will be calculated by reference to the exchange parity with one of the fiat currencies established at that date by one of the *exchange* platforms explained earlier in the paper.

### B. Legislative implications

Given the legal nature of these assets and the fact that possession of the keys and wallet number automatically equates to ownership, another question to consider is: what would be the need to transfer these cryptocurrencies by inheritance, as long as they can be passed on directly? We believe that in all cases, in order to take possession of, for example, a bank safe deposit box or other *wallet* transfer options, including access to the exchange platforms on which these currencies are held, heirs will have to prove their status, in accordance with the laws of each country. At the same time, the act, *as instrumentum*, of the inheritance of cryptocurrencies, such as the certificate of inheritance, will represent the title deed for the heirs and thus they will be able to prove the source of the inherited assets. But this question also raises a number of new topical issues, such as: how did the author acquire these coins? What is the means of proof, of evidence of their acquisition? Or, because they are movable goods, will the simple fact of possessing them be considered as ownership? What will happen if these cryptocurrencies have not been legally owned?

The answers to these questions have not yet been given in case law and literature, and in the coming years we will carefully monitor the legal and doctrinal evolution of the subject in order to fully regulate and clarify the complex aspects of the transmission of cryptocurrencies by succession.

## V. CONCLUSIONS

The transfer of cryptocurrencies through succession is a subject of study that requires an inter- and multidisciplinary approach, which must include both technical and legal or fiscal aspects. Awareness of the importance of succession planning regarding this relatively recent asset type is essential for digital currency users. Thus, it is necessary to implement appropriate measures to ensure the safe transfer of these digital assets to the heirs. Legal systems must evolve to integrate new types of assets and to ensure the protection of the rights of successors. The present study wishes to present some specificities of digital currencies and to offer a series of recommendations for improving the existing legislation and for the implementation of effective technical solutions, thus contributing to the development of an appropriate legal framework for legacies with cryptocurrencies as their object.